\documentstyle[12pt,a4]{article}
\begin{document}
\baselineskip 18pt
\def\today{\ifcase\month\or
 January\or February\or March\or April\or May\or June\or
 July\or August\or September\or October\or November\or December\fi
 \space\number\day, \number\year}
\def\thebibliography#1{\section*{References\markboth
 {References}{References}}\list
 {[\arabic{enumi}]}{\settowidth\labelwidth{[#1]}
 \leftmargin\labelwidth
 \advance\leftmargin\labelsep
 \usecounter{enumi}}
 \def\newblock{\hskip .11em plus .33em minus .07em}
 \sloppy
 \sfcode`\.=1000\relax}
\let\endthebibliography=\endlist
%%%%%%%%%%%%%%%%%%%%%%%%%%%%%%%%%%%%%%%%%%%%%%
%%%%%%%                                %%%%%%%
%%%%%%%	        Journal code           %%%%%%%
%%%%%%%                                %%%%%%%
%%%%%%%%%%%%%%%%%%%%%%%%%%%%%%%%%%%%%%%%%%%%%%
\def\MPLA#1#2#3{Mod. Phys. Lett. {\bf A#1} (#2) #3}
\def\PRD#1#2#3{Phys. Rev. {\bf D#1} (#2) #3}
\def\NPB#1#2#3{Nucl. Phys. {\bf B#1} (#2) #3}
\def\PTP#1#2#3{Prog. Theor. Phys. {\bf #1} (#2) #3}
\def\ZPC#1#2#3{Z. Phys. {\bf C#1} (#2) #3}
\def\EPJC#1#2#3{Eur. Phys. J. {\bf C#1} (#2) #3}
\def\PLB#1#2#3{Phys. Lett. {\bf B#1} (#2) #3}
\def\PRL#1#2#3{Phys. Rev. Lett. {\bf #1} (#2) #3}
\def\PRep#1#2#3{Phys. Rep. {\bf #1} (#2) #3}
\def\RMP#1#2#3{Rev. Mod. Phys. {\bf #1} (#2) #3}
\begin{titlepage}
\begin{flushright}
\begin{tabular}{l}
OCHA-PP-181
 \\
\end{tabular}
\end{flushright}
\vskip 1.0 true cm 
\begin{center}
{\large {\bf Neutron Electric Dipole Moment and \\
 Extension of the Standard Model}}
\footnote{
To appear in the Proceedings of RCNP Workshop on Production  
of Ultra Cold Neutrons and Fundamental Physics.
}
\vskip 2.0 true cm
\renewcommand{\thefootnote}
{\fnsymbol{footnote}}
Noriyuki Oshimo 
\\
\vskip 0.5 true cm 
{\it Department of Physics} \\ 
{\it Ochanomizu University}  \\
{\it Bunkyo-ku, Tokyo 112-8610, Japan}  \\
\end{center}

\vskip 3.0 true cm

\centerline{\bf Abstract}
\medskip
     A nonvanishing value for the electric dipole moment (EDM) 
of the neutron is a prominent signature for $CP$ violation.  
The EDM induced by the Kobayashi-Maskawa mechanism of 
the standard model (SM) has a small magnitude and its detection 
will be very difficult.  
However, since baryon asymmetry of the universe cannot be 
accounted for by the SM, there should exist some other source 
of $CP$ violation, which may generate a large magnitude for the EDM.  
One of the most hopeful candidates for physics beyond the SM is 
the supersymmetric standard model, which contains such 
sources of $CP$ violation.  
This model suggests that the EDM has a magnitude not much smaller than 
the present experimental bounds.  
Progress in measuring the EDM 
provides very interesting information about extension of the SM.  

\vspace{1 cm}

\medskip

\end{titlepage}

\newpage 
\setcounter{footnote}{0}
 
     Nature does not respect invariance for $CP$ transformation, 
although only a few demonstrations are known.  
To date, in the $K$ meson system, 
and very recently in the $B$ meson system, 
$CP$-violating phenomena have been observed.  
Another manifestation of $CP$ violation is baryon 
asymmetry of the universe.  
Within the framework of particle physics, the former could be described 
well by the Kobayashi-Maskawa mechanism of the standard model (SM).  
However, the latter cannot be explained by the SM.  

     A nonvanishing value of the electric dipole moment (EDM) of the 
neutron, though not yet found, unambiguously represents $CP$ violation.  
Experimental measurements have put its upper bounds as 
$|d_n|<6.3\times 10^{-26} e$cm.  
In the SM, disregarding the strong $CP$ problem, 
the EDM is generated only at the three-loop level.  
Its predicted magnitude is at most of order $10^{-31}e$cm,  
which is extremely smaller than the present experimental bounds.  
However, a new source of $CP$ violation could easily enhance the EDM.  
In fact, from various points of view, the SM is considered to be an effective 
theory of a more fundamental one for the energy scale of 100 GeV$-$1 TeV.  
Above this energy scale, some new physics may emerge, 
making extension of the SM inevitable.   
Such an extension involves mostly new sources of $CP$ violation.  
These sources may serve for baryogenesis, and may cause 
new $CP$-violating phenomena which are testable in experiments.  
Studying the EDM, both theoretically and experimentally, provides very useful information about physics beyond the SM.  

     One of the most hopeful candidates for the extension of the SM 
is the supersymmetric standard model (SSM).  
Supersymmetry is a symmetry between the particles with 
different spin quantum numbers.  
Therefore, the SSM 
involves new particles which are supersymmetric partners of 
the ordinary particles contained in the SM.  
Through this model, the SM can be consistently embedded in the 
scheme of grand unification, the unification of the 
SM gauge groups into a single gauge group.  
In addition, the SSM contains several new sources of $CP$ violation, 
which could generate baryon asymmetry of the universe.   
Examination of the SSM thus may be performed by investigating 
$CP$ violation intrinsic in this model.  

     A prominent effect of the new sources of $CP$ violation 
in the SSM is found in the EDM of the neutron \cite{kizukuri}.  
Its value could be around the experimental bounds.   
The reason of such a large magnitude is that the EDM is generated 
at the one-loop level.  
The EDM of a quark $q$ arises from three types of one-loop 
diagrams in which supersymmetric particles are exchanged:  
(i.a) charginos $\omega$ and squarks $\tilde q'$, 
(i.b) neutralinos $\chi$ and squarks $\tilde q$,  
(i.c) gluinos $\tilde g$ and squarks $\tilde q$.  
The charginos and the neutralinos are supersymmetric partners of the 
SU(2)$\times$U(1) gauge bosons and the Higgs bosons, with 
their electric charges being $-1$ and $0$, respectively.  
The gluinos are supersymmetric partners of the SU(3) gauge bosons.  
The squarks are supersymmetric partners of the quarks.  
In the diagram (i.a) the third components of weak isospin for 
the quark and the squarks are different, 
while in the diagrams (i.b) and (i.c) they are the same.  
From the nonrelativistic quark model, the neutron EDM is given by 
$d_n=(4d_d-d_u)/3$, where $d_d$ and $d_u$ respectively denote 
the EDMs of the down quark and the up quark.  

     The neutron EDM is also generated 
by two-loop diagrams \cite{kadoyoshi}.  
In the SSM, a $CP$-odd three-body coupling for the $W$ bosons and 
the photon arises from three-types of one-loop diagrams mediated by 
supersymmetric particles:  (ii.a) charginos and neutralinos, 
(ii.b) squarks, (ii.c) sleptons.  
The sleptons are supersymmetric partners of the leptons.  
Through one-loop diagrams generated by the SM interactions, 
the $CP$-odd $WW\gamma$ coupling leads to the EDMs of 
the quarks, which are at the two-loop level.  

     The interactions relevant to generation of the neutron EDM 
are described by the model parameters of the SSM contained in 
the mass terms of supersymmetric particles.  
The mass matrices of the charginos and the neutralinos consist 
of the SU(2) gaugino mass $m_2$, the U(1) gaugino mass $m_1$, 
the higgsino mass $m_H$, and the ratio $\tan\beta$ of 
the vacuum expectation values for the Higgs bosons.  
The gluino mass is given by the SU(3) gaugino mass $m_3$.  
The mass-squared matrices of the squarks or sleptons consist of the 
gravitino mass $m_{3/2}$, the left-handed and right-handed squark or 
slepton masses $M_L$ and $M_R$, the coefficient $A$ for the trilinear couplings 
of the Higgs bosons and the squarks or sleptons, 
in addition to $m_H$ and $\tan\beta$.  
Under the scheme of grand unification, the gaugino masses are related 
to each other as $3m_1/5g_1^2=m_2/g_2^2=m_3/g_3^2$.  
Assuming the ordinary mechanism of supersymmetry 
breaking by the model based on $N=1$ supergravity, 
the relation $m_{3/2}\approx M_L\approx M_R$ holds.  

     The magnitude of the neutron EDM correlates with 
the $CP$-violating phases and the masses of intermediate particles.  
By redefining the phases of particle fields, without loss of generality, 
we can take the SSM parameters except $m_H$ and $A$ as real and positive.  
The complex phases of $m_H$ and $A$, which are expressed as 
$m_H=|m_H|\exp(i\theta)$ and $A=|A|\exp(i\alpha)$, 
are physical and thus induce $CP$ violation.  
The masses of the charginos and the neutralinos are of order of 
$m_2$ and $m_H$, and the gluino mass is about three times $m_2$.  
The masses of the squarks and the slepton are of order of $m_{3/2}$.  
Since the supersymmetric particles have not yet been discovered in 
collider experiments, their masses should be larger than 100 GeV roughly.  
The lightest supersymmetric particle should be the lightest neutralino 
from a cosmological consideration, which is realized for 
$m_2, |m_H|<m_{3/2}$.  
In general, the EDM becomes suppressed as the complex phases 
$\theta$, $\alpha$ decrease, 
or the mass parameters $m_2$, $|m_H|$, $m_{3/2}$ increase.  

     Assuming unsuppressed complex phases, for 
$1\ {\rm TeV}<m_{3/2}<10$ TeV and $100\ {\rm GeV}<m_2, |m_H|<1$ TeV, 
the contributions from the one-loop diagrams give the neutron EDM 
a value $d_n\sim (10^{-26}-10^{-24}) e$cm.  
Certain regions of the parameter space can already be excluded by 
the experimental bounds for the EDM available at present.  
If the three mass parameters are all smaller than 1 TeV, 
the predicted magnitude for the EDM becomes too large.  
Among the three types of one-loop diagrams, the contribution 
of the diagram (i.a) is generally dominant.  
This contribution increases as the value of $\tan\beta$ becomes large.  
The two-loop contribution coming from the diagram (ii.a) gives 
$d_n\sim (10^{-28}-10^{-26}) e$cm for $100\ {\rm GeV}<m_2, |m_H|<1$ TeV.  
The dependency of this contribution on $\tan\beta$ is opposite 
to the above one-loop contribution.  
The other two-loop contributions from the diagrams (ii.b) and (ii.c) 
yield smaller magnitudes as long as the inequality 
$m_2, |m_H|<m_{3/2}$ is satisfied.  
 
     The predicted magnitude of the neutron EDM becomes small as 
the value of $m_{3/2}$ increases, 
since all the one-loop contributions involve the squarks.  
On the other hand, the two-loop contribution via the diagram (ii.a) 
is independent of the squarks.   
If the squarks are very heavy, 
the two-loop contribution could become dominant.    
For instance, for $m_{3/2}\sim 10$ TeV with $m_2, |m_H|\sim 100$ GeV  
and $\tan\beta\sim 1$, the one-loop contributions become comparable 
to or smaller than the two-loop contribution.  
Therefore, irrespectively of $m_{3/2}$, 
the EDM is expected to have a magnitude $(10^{-28}-10^{-25}) e$cm 
for $100\ {\rm GeV}<m_2, |m_H|<1$ TeV, 

     The EDM of the neutron may also have a small magnitude, if 
the $CP$-violating phases intrinsic in the SSM are suppressed.  
However, such a situation would require reasonable explanation 
for the reason why the SSM parameters are almost real whereas 
the quark Yukawa coupling constants have unsuppressed complex phases.  
Furthermore, the magnitudes of the complex phases are bounded 
from below, if baryon asymmetry of the universe should be generated.  

     Within the framework of the SSM, baryogenesis could occur 
at the electroweak phase transition of the early universe \cite{aoki}.  
Provided that the phase transition is strongly first order, 
the bubbles of broken gauge symmetry nucleate in the 
SU(2)$\times$U(1) symmetric phase.  
In the course of expansion of the bubbles, 
particles are reflected or transmitted by the bubble walls.  
Large $CP$-violating effects are then caused by 
the interactions of the charginos or the top squarks with the wall.  
Between the $CP$ conjugate states of these particles, 
the reflection and transmission rates become asymmetric, 
leading to net hypercharge flux emitted into the symmetric phase.  
The resultant nonvanishing hypercharge density puts a bias to  
the equilibrium conditions for nonvanishing baryon number density, 
which is achieved through electroweak anomaly.  
The yielded net baryon number is subsequently captured by the phase 
of broken gauge symmetry, as this phase expands.  
After the phase transition, baryon number of the universe is 
not changed, since baryon number violation by electroweak anomaly 
becomes negligible.  

     The observed ratio of baryon number to entropy in the universe 
can be produced, if $CP$-violating effects are large and certain 
conditions for the electroweak phase transition are satisfied.  
For the chargino interactions to properly accomplish baryogenesis,  
the parameters should be within the ranges 
$|\sin\theta|>0.1$ and $100\ {\rm GeV}<m_2, |m_H|<1$ TeV.  
The neutron EDM is then expected to be $(10^{-27}-10^{-25}) e$cm 
for $1\ {\rm TeV}<m_{3/2}<10$ TeV.  
Baryogenesis by the top squark interactions gives the ranges 
$\sin\alpha\sim 1$, $|\sin\theta|\ll 1$, and $100\ {\rm GeV}<m_{2/3}<1$ TeV.  
If the phase $\theta$ is sufficiently suppressed, the EDM receives 
a dominant contribution from the diagram (i.c).  
For $100\ {\rm GeV}<m_2, |m_H|<1$ TeV, the EDM is expected to 
have a magnitude $(10^{-26}-10^{-25}) e$cm.  
Hence, baryon asymmetry of the universe suggests that the EDM 
is larger than of order of $10^{-27} e$cm.  

     The new sources of $CP$ violation in the SSM also generate 
the EDM of the electron at both the one-loop and the two-loop levels.  
The one-loop diagrams are mediated by the charginos and the 
neutralinos together with the sleptons.  
The two-loop diagrams originate in the $CP$-odd $WW\gamma$ 
coupling.  
Its predicted magnitude is roughly $0.1-0.01$ times 
that of the neutron EDM, which is far larger than the SM prediction 
and could be explored in experiments.  

     The mechanism of generating the neutron EDM in the SSM is related 
with the radiative decay of the bottom quark \cite{oshimo}.  
The bottom quark can decay into the strange quark and the 
photon through one-loop diagrams mediated by supersymmetric particles.  
Some information on $CP$ violation relevant to the neutron EDM 
could be obtained by collider experiments.  
For instance, a $CP$-violating effect is found in 
the decay rate asymmetry between the $CP$ conjugate processes 
$b\to s\gamma$ and $\bar b\to\bar s\gamma$ \cite{cho}.  
For the parameters ranges $\sin\alpha\sim 1$, 
$|\sin\theta|\ll 1$, and $100\ {\rm GeV}<m_2, |m_H|, m_{2/3}<1$ TeV, 
the decay rate asymmetry can become larger than the SM prediction, 
and may be detectable in the near future.  

     The EDM of the neutron provides interesting and useful  
information for particle physics.  
Although its predicted magnitude by the SM is extremely small, 
a much larger magnitude could be generated by new sources of $CP$ violation, 
which are contained in various extensions of the SM.   
Baryon asymmetry of the universe, which cannot be explained by the SM, 
also suggests the existence of a new source of $CP$ violation.  
The SSM, which is one of the most plausible extensions of the SM, 
predicts that the magnitude of the EDM is not much smaller than 
the present experimental bounds in wide ranges of reasonable 
parameter values.  
Any refinement of the experimental bounds is thus useful for studying  
physics beyond the SM.  
Measuring the EDM is a very important subject for 
experiments which use the ultra cold neutron.

\smallskip 
        
     This article is based on the works in collaboration with 
M. Aoki, G.C. Cho, T. Kadoyoshi, Y. Kizukuri, and A. Sugamoto.  

\vskip 3.0 true cm


\newpage

\begin{thebibliography}{99}
\bibitem{kizukuri}
    Y. Kizukuri and N. Oshimo, 
    \PRD{45}{1992}{1806}; {\bf D46} (1992) 3025.   
\bibitem{kadoyoshi}
    T. Kadoyoshi and N. Oshimo, 
    \PRD{55}{1997}{1481}.   
\bibitem{aoki}
    M. Aoki, N. Oshimo, and A. Sugamoto,  
    \PTP{98}{1997}{1179}; {\bf 98} (1997) 1325.   
\bibitem{oshimo}  
 	N. Oshimo, \NPB{404}{1993}{20}.  
\bibitem{cho}
    M. Aoki, G.C. Cho, and N. Oshimo,
    \PRD{60}{1999}{035004}; \NPB{554}{1999}{50}. 

\end{thebibliography}
\end{document}